\documentclass[%
 aip,
 jmp,%
 amsmath,amssymb,
 reprint,%
]{revtex4-1}

\usepackage{graphicx}
\usepackage{dcolumn}
\usepackage{bm}

\begin{document}

\preprint{AIP/123-QED}

\title[]{Magnetic Flux Dynamics in Horizontally Cooled Superconducting Cavities}

\author{M. Martinello}\email{mmartine@fnal.gov.}%
 \affiliation{Fermi National Accelerator Laboratory, Batavia, Illinois 60510, USA.}%
 \affiliation{Department of Physics, Illinois Institute of Technology, Chicago, Illinois 60616, USA.}
\author{M. Checchin}%
 \affiliation{Fermi National Accelerator Laboratory, Batavia, Illinois 60510, USA.}%
 \affiliation{Department of Physics, Illinois Institute of Technology, Chicago, Illinois 60616, USA.}
\author{A. Grassellino}\email{annag@fnal.gov.}%
 \affiliation{Fermi National Accelerator Laboratory, Batavia, Illinois 60510, USA.}%
\author{A.C. Crawford}%
 \affiliation{Fermi National Accelerator Laboratory, Batavia, Illinois 60510, USA.}%
\author{O. Melnychuk}%
 \affiliation{Fermi National Accelerator Laboratory, Batavia, Illinois 60510, USA.}%
\author{A. Romanenko}%
 \affiliation{Fermi National Accelerator Laboratory, Batavia, Illinois 60510, USA.}%
\author{D.A. Sergatskov}%
 \affiliation{Fermi National Accelerator Laboratory, Batavia, Illinois 60510, USA.}%

\date{\today}

\begin{abstract}
Previous studies on magnetic flux expulsion as a function of cooling details have been performed for superconducting niobium cavities with the cavity beam axis placed parallel respect to the helium cooling flow, and findings showed that for sufficient cooling thermogradients all magnetic flux could be expelled and very low residual resistance could be achieved. In this paper we investigate the flux trapping and its impact on radio frequency surface resistance when the resonators are positioned perpendicularly to the helium cooling flow, which is representative of how superconducting radio-frequency (SRF) cavities are cooled in an accelerator. We also extend the studies to different directions of applied magnetic field surrounding the resonator. Results show that in the cavity horizontal configuration there is a different impact of the various field components on the final surface resistance, and that several parameters have to be considered to understand flux dynamics. A newly discovered phenomenon of concentration of flux lines at the cavity top leading to cavity equator temperature rise is presented.
\end{abstract}

\maketitle

%

\section{\label{sec:level1}Introduction}
Trapped magnetic flux in superconducting resonators contributes to the radio-frequency (RF) surface resistance ($R_{s}$), in the form of the temperature-independent residual resistance $R_{0}$\cite{1}.\\
\indent This residual resistance due to trapped flux plays an important role in the SRF cavity performance, potentially degrading the efficiency of the cavity. Recent studies\cite{2,3,4} have shown that performing fast cool-downs, with high thermogradients along the cavity, is vital to obtain efficient ambient magnetic flux expulsion, and that slow and homogeneous cooling through transition leads to full flux trapping. As an example, using the fast cooling technique, residual resistances values as low as $1$ n$\Omega$ have been obtained with $1.3$ GHz nitrogen doped niobium cavities\cite{5} in up to $20$ mG magnetic fields, and even $5$ n$\Omega$ in $190$ mG, compared to $15$ n$\Omega$ in $5$ mG for slow cooling through critical temperature $T_{c}$\cite{3}. These examples elucidate how the cool-down regime is a crucial parameter to maximize superconducting accelerating cavities efficiency.\\ 
Several continuous wave (CW) accelerators currently being built worldwide  -as x-ray FELs like LCLS-II at SLAC\cite{6,7}- require very high quality factors (highly efficient SRF cavities) to reduce cryogenic costs. Therefore, the understanding of the cool-down dynamics in a configuration that resembles the cavity in an accelerator, in presence of magnetic field levels comparable to those present in shielded cavities placed in a cryomodule, is crucial in order to investigate how to minimize the surface losses due to trapped flux.\\
\indent The goal of the work presented in this paper is therefore studying the impact of different external magnetic field orientations on the residual resistance when the cavity is transverse respect to the cooling direction, exactly as it happens when cavities are placed in an accelerator. In particular, we focused on the difference introduced by an external magnetic field applied axially versus orthogonally to the cavity axis.
\section{\label{sec:level2}EXPERIMENTAL SET-UP}
\begin{figure}[t]
\centering
\includegraphics[scale=0.8]{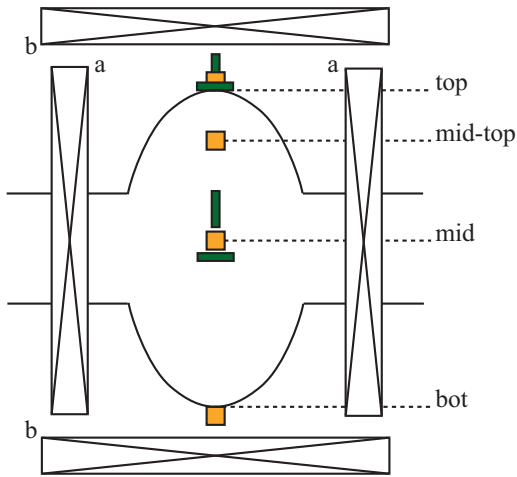}
\caption{Horizontal cool-down cavity set-up.}
\label{fig:Set}
\end{figure}

In this study we used a single cell $1.3$ GHz TESLA type nitrogen doped niobium cavity, the same as used in previous work\cite{3}. It is worth mentioning that this cavity is currently the highest quality factor cavity ever measured with a $Q_{0}>1\cdot 10^{11}$ up to the highest fields $30$ MV/m at $1.5$ K and with a $Q_{0}>5\cdot 10^{10}$ up to $30$ MV/m at $2$ K.\\
\indent The set-up with the cavity placed horizontally in the cryostat is shown in FIG. \ref{fig:Set}. Two Helmholtz coils were placed orthogonal one to each other (FIG. \ref{fig:Set}), one parallel to the cavity axis (a) and the other perpendicular to it (b). In addition, four Cernox thermometers were placed on the cavity equator (orange squares in FIG. \ref{fig:Set}): one on the very bottom of the cell, one in the middle, one on the very top of the cell, and one half way in between of the top and the middle ones.\\ 
\indent The external magnetic field applied to the cavity was measured by means of four single-axis Bartington Mag-01H cryogenic fluxgate magnetometers (green rectangles in FIG. \ref{fig:Set}). Two of them were placed vertically, one at the very top of the cell and one at the middle, while the other two were placed at the same positions but horizontally.\\ 
\indent Several fast cool-downs were performed under different magnetic field orientations (orthogonal or axial) and same magnitude, about $10$ mG. In order to obtain different thermogradients across the cavity, different starting temperatures were chosen for these fast cool-downs.
\section{\label{sec:level3}CAVITY COOL-DOWN DYNAMICS}
In order to better understand how the cavity RF surface resistance is affected by different cool-downs, we start by discussing the dynamics with which the superconducting transition takes place along the cavity.\\
\indent When the cool-down is performed with the cavity oriented horizontally with respect the cryostat axis, the boundary between the superconducting (SC) and the normal conducting (NC) phases will move from the very bottom to the very top point of the cell equator, rather than from beam tube to beam tube as in previous studies\cite{2,3}. This can cause significant differences from the vertical geometry, since now the final escaping place for flux is the equator, most important area for RF losses.\\
\begin{figure*}[!]
  \includegraphics[scale=0.9]{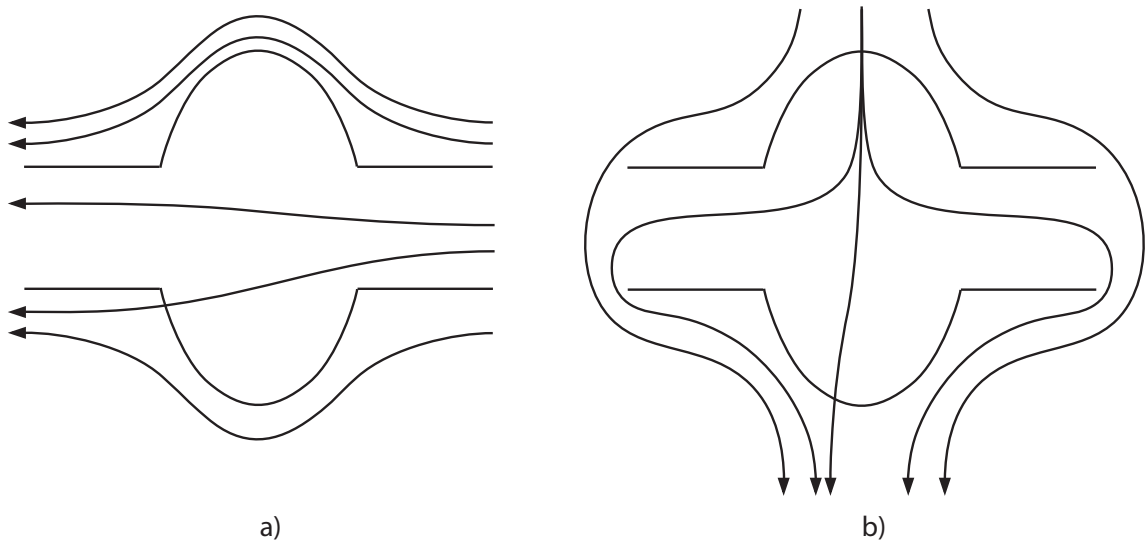}
  \caption{Field redistribution in the Meissner state with magnetic field applied a) axially and b) orthogonally.}
  \label{fig:AxOrt}
\end{figure*}
\begin{table*}[!]
\begin{tabular}{ccccc}
\hline
\\
\hspace{0.3cm} Name\hspace{0.3cm} &\hspace{0.3cm} Field Magnitude\hspace{0.3cm} &\hspace{0.3cm} Field Orientation\hspace{0.3cm} &\hspace{0.3cm} Start Temperature\hspace{0.3cm} &\hspace{0.3cm} $R_{0}$ @ $16$ MV/m\hspace{0.3cm} \\
 & [mG] & & [K] & [n$\Omega$]\\[6pt]
\hline
\\
 1Ax & 10 & Axial & 300 & 2.1\\[6pt]
 2Ax & 10 & Axial & 50 & 5.3\\[6pt]
 3Ax & 10 & Axial & 300 & 2.9\\[6pt]
\hline
\\
 1Ort & 10 & Orthogonal & 300 & 6.3\\[6pt]
 2Ort & 10 & Orthogonal & 260 & 6.1\\[6pt]
 3Ort & 10 & Orthogonal & 170 & 7.7\\[6pt]
 4Ort & 10 & Orthogonal & 25 & 13.9\\[6pt]
\hline
\end{tabular}
\caption{Cool-downs summary and associated measured residual resistance.}
\label{table:table}
\end{table*}
\indent Let us consider the two different cases of field to be expelled oriented axially or orthogonally. When a magnetic field is applied axially to the cavity during the transition, with sufficient thermogradients this field will be expelled from the cavity walls because of the Meissner effect, as depicted in FIG. \ref{fig:AxOrt}a. In the Meissner state the magnetic field can be confined outside the cavity volume, or it can pass inside the cavity through the beam pipe. Because of the axial direction of the field, the expulsion can be efficient because the flux lines that cross the cavity walls always have an easy path to follow during the transition. If the flux expulsion is not efficient, it may happen that some flux lines remains pinned, crossing the cavity walls, and increasing the losses.\\
\begin{figure}[!]
\centering
\includegraphics[scale=1]{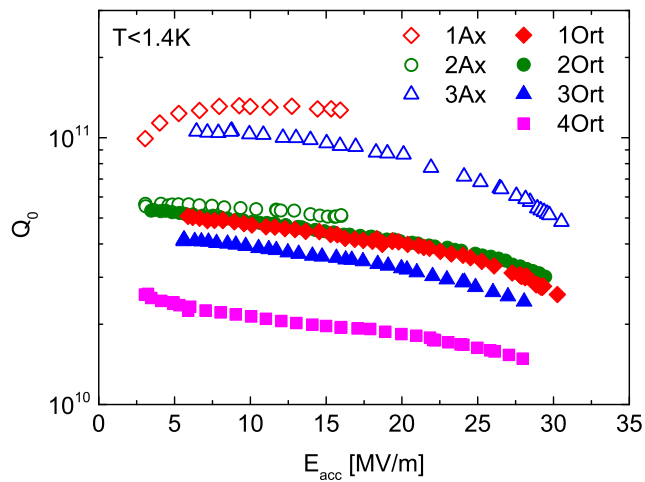}
\caption{$Q_{0}$ versus accelerating field measured at $T<1.4$ K.}
\label{fig:QovsEacc}
\end{figure}
\begin{figure*}[!]
\includegraphics[scale=1]{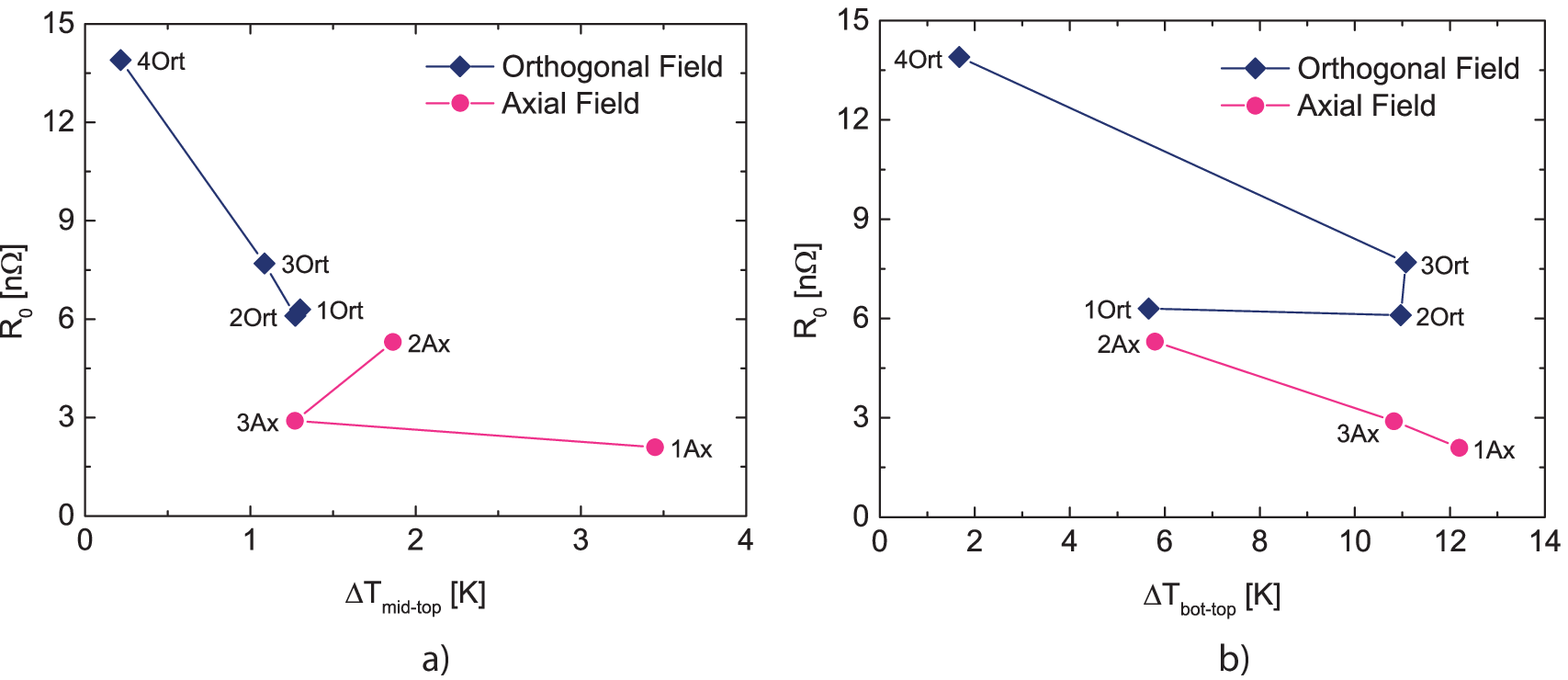}
\caption{Residual resistance versus mid-top a) and bot-top b) thermogradients.}
\label{fig:DT}
\end{figure*}
\indent When the applied field is perpendicular to the cavity axis during the SC transition, the dynamics will be different: first magnetic field lines will be bent because of the Meissner effect, and then redistributed in three possible different ways: i) completely outside the cavity, ii) escaping through the beam pipes, or iii) escaping across the cavity wall if pinning centres are present, when a non-efficient expulsion occurs (FIG. \ref{fig:AxOrt}b). Because of the orthogonal field orientation, the magnetic flux lines redistributed with the ii) and iii) mechanisms, do not have any possibility to escape from the cavity inner volume except crossing the cavity walls. Considering a sharp SC-NC interface, these flux lines will be concentrated in the normal conducting region, that will become smaller and smaller as the transition boundary advances, till they will be squeezed at the very top point of the cell equator. Indeed, this point of the cell will be the last cooled below the transition temperature $T_{c}$, becoming a "flux hole" in the superconductor, from which it is energetically not favourable for flux to escape as it the only way out would be via crossing already superconducting regions.\\
\indent The situation depicted implies that the geometry of the system could lead to incomplete Meissner effect, even though the thermal gradient across the cavity is high enough to provide efficient flux expulsion for the axial configuration.
\section{\label{sec:level4}DATA ANALYSIS}
\begin{figure}[!]
\centering
\includegraphics[scale=1]{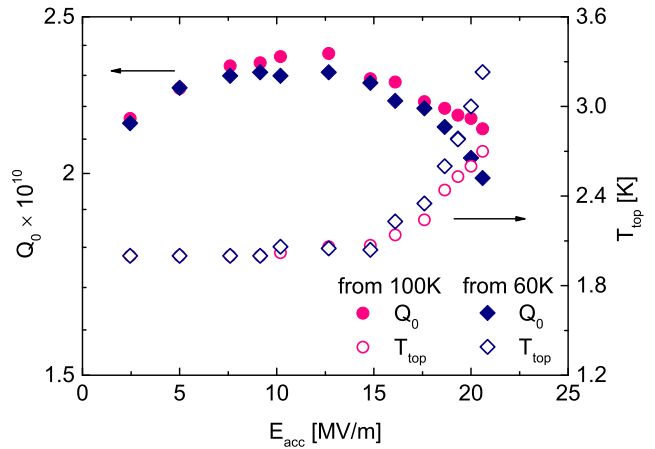}
\caption{$Q_{0}$ and $T_{top}$ versus accelerating field for HTS measurements done with cool-downs started at $100$ K and $60$ K.}
\label{fig:HTS}
\end{figure}

The cavity RF measurements were performed at the Fermilab SRF cavity vertical test facility. The unloaded Q-factor ($Q_{0}$) versus accelerating field ($E_{acc}$) curves were acquired at $2$ K and at the lowest temperature achievable, that is lower than the calibration range of the thermometers ($T<1.4$ K). At such low temperatures the cavity performance are dominated by the temperature-independent part (residual resistance $R_{0}$), therefore the $Q_{0}$ versus accelerating field curve is not affected by temperature variations.\\
\indent The $Q_{0}$ versus accelerating field curves acquired at $T<1.4$ K are shown in FIG. \ref{fig:QovsEacc}, while the cool down conditions of the data series are summarized in TABLE \ref{table:table}.\\
\indent The cool-downs of the data series named nAx (axial) were performed applying $10$ mG of external field parallel to the cavity axis.\\ 
\indent The highest quality factor was reached with 1Ax, the Q-factor increases slightly at low field and it reaches $1.3\cdot 10^{11}$ at $16$ MV/m. Lower performance are shown by 3Ax curve. It decreases considerably with the accelerating field, showing the typical slope due to trapped magnetic flux\cite{8,9}. The worst performance for the axial series are shown by 2Ax curve, in which the Q-factor reaches only $5.1\cdot 10^{10}$ at $16$ MV/m.\\ 
\indent The cool-downs of the nOrt (orthogonal) series were performed applying $10$ mG orthogonally to the cavity axis. First thing that stands out is that all the curves of the orthogonal series show worst performance than the axial series, and they all show $Q_{0}$ degradation with the field due to trapped flux. The best performance for the orthogonal series are shown by 2Ort with $Q_{0}=4.3\cdot 10^{10}$ at $16$ MV/m, while the lowest Q-factor values are given by the 4Ort curve, and in this case the Q-factor is $1.9\cdot 10^{10}$ at $16$ MV/m.\\
\indent The residual resistance at $16$ MV/m was calculated as $G/Q_{0}$ ($G=270$ $\Omega$), since, as already mentioned, the surface resistance at $T<1.4$ K is affected only by the temperature independent part. In TABLE \ref{table:table} the residual resistance of each series is reported.\\
\indent Considering the data series axial and orthogonal individually, different cool-downs lead to different residual resistances, as already found for the usual vertical configuration\cite{1}.\\
\indent We suggest that two useful parameters to describe the dynamics of the cavity cool-down are the thermo-gradients $\Delta T_{bot-top}$ and $\Delta T_{mid-top}$. The first one corresponds to the temperature difference between the top and the bottom of the cell, when the bottom reaches $T_{c}$. The second one is the temperature difference between the top and the mid position of the cell when mid pass through the SC transition.\\
\begin{figure}[!]
\centering
\includegraphics[scale=1]{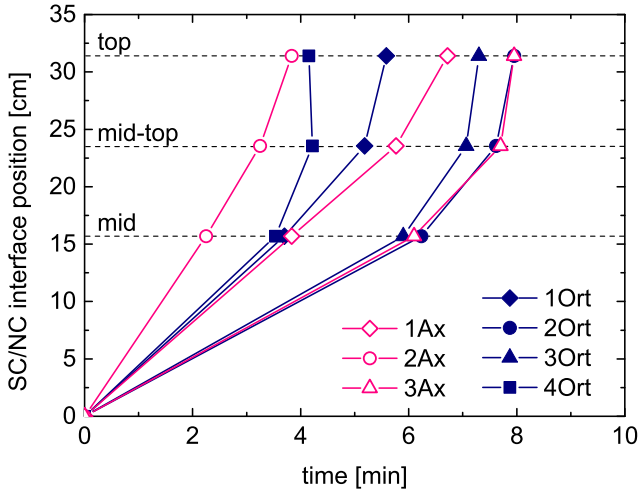}
\caption{\label{fig:Cool}Evolution of the SC/NC interface during the cool-down.}
\end{figure}
\indent The residual resistance as function of the thermogradients $\Delta T_{mid-top}$ and $\Delta T_{bot-top}$ is shown in FIG. \ref{fig:DT}. Looking at FIG. \ref{fig:DT}a, the residual resistance, for the orthogonal series, seems to follow a trend with $\Delta T_{mid-top}$ thermogradient. No particular trend appears for the axial series. Conversely, the linear trend of the residual resistance as function of the thermogradient $\Delta T_{bot-top}$ appears only for the axial series (FIG. \ref{fig:DT}b).\\
\indent Comparing 2Ort and 3Ort, they show the same $\Delta T_{bot-top}$ thermogradients, but lower $R_{0}$ value is measured for 3Ort, which instead has a higher $\Delta T_{mid-top}$.\\
\indent Therefore, the data suggests that both mid-top and bottom-top thermogradients may play an important role to determine the residual losses, as one could intuitively expect: cooling details may vary as the SC-NC boundary progresses along the cavity profile, \textit{but what matters for efficient flux expulsion is thermogradients at SC-NC phase front}.\\
\indent In order to better understand the global behaviour of the cavity during the cool-down, we therefore investigated also the SC-NC interface evolution during the cool-down. Setting to zero the time at which the cavity bottom passes through transition, the position of the SC-NC interface can be plotted against the time it takes moving from one thermometer position to another, as shown in FIG. \ref{fig:Cool}.\\
\indent It is important to underline that the slope of the segment connecting two points corresponds to the average speed (cm/min) of the SC-NC interface along the cell, that should not be confused with the cooling speed (K/min) which, on the contrary, seems not to be a key parameter for cavity losses.\\
\indent One of the important things that we can conclude from FIG. \ref{fig:Cool} is that indeed thermogradients per unit length are not constant throughout the movement of the SC-NC interface, but that they tend to narrow down as the boundary moves towards the top. This is perhaps an effect of the cavity starting out warm but then rapidly cooling by conduction. This could potentially cause more flux to get trapped at the top, which in the horizontal cavity case is an equator, causing a certain performance degradation compared to the vertical cavity orientation.\\
\begin{figure*}[!]
\includegraphics[scale=1]{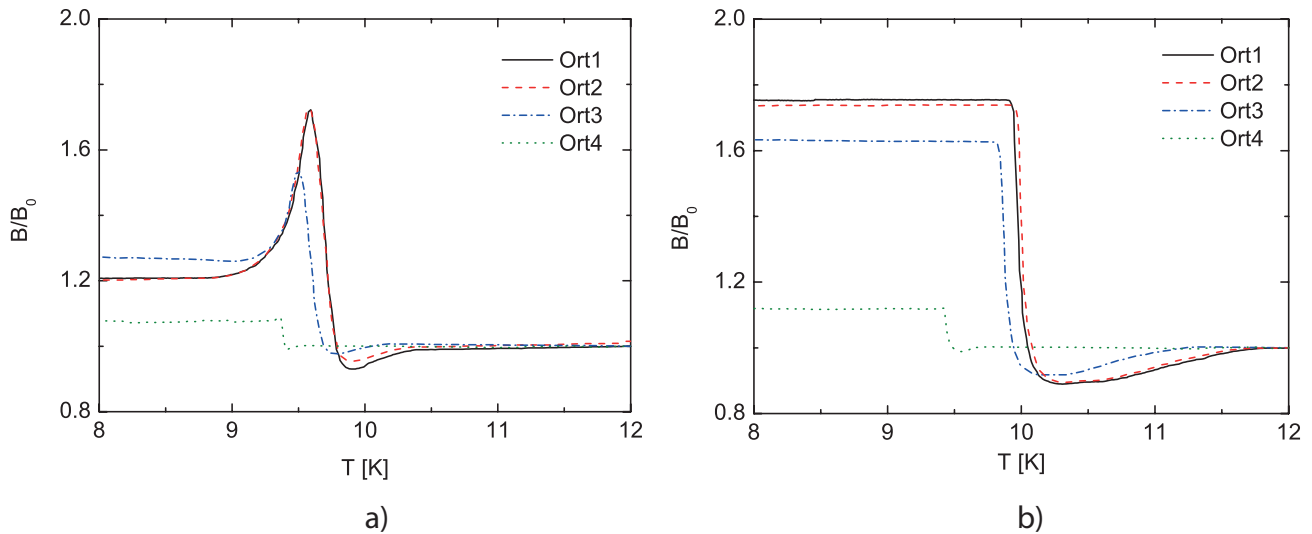}
\caption{Relative orthogonal magnetic field for the orthogonal applied magnetic field cool-downs at a) the cavity mid position and b) on top.}
\label{fig:BB0}
\end{figure*}
\indent As an extreme case, it can be noticed that for 4Ort the top of the cavity passes through transition before the mid-top position. Then, the last point which becomes SC is not the very top −as in all the other cases− implying that the magnetic flux not escaped from the cavity internal volume will not be squeezed at the very top of the cavity, but redistributed in the nearby zone. The cooling scenario will now be better described by a superconducting phase nucleation dynamics rather than by a sharp SC-NC interface movement across the cavity, as suggested by the fact that mid-top position becomes SC after the very top of the cavity. Because of the superconducting phase nucleation dynamics, the incomplete Meissner effect is enhanced by the presence of normal conducting islands surrounded by the SC material, leading to a not efficient flux expulsion and large residual losses, as already described in previous work\cite{2}.\\
\begin{figure}[b]
\centering
\includegraphics[scale=0.9]{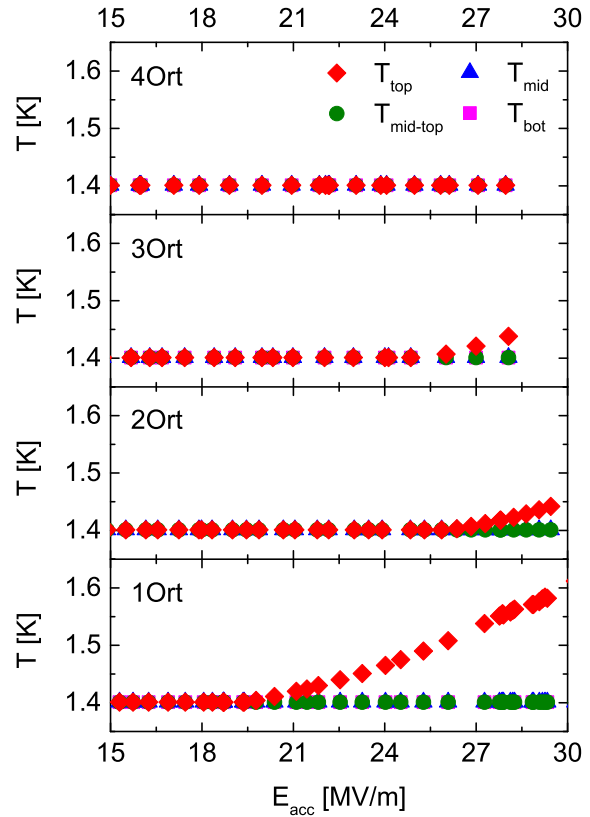}
\caption{Temperature variation versus the accelerating field.}
\label{fig:Temp}
\end{figure}
\indent The magnetic field data shown in FIG. \ref{fig:BB0} were acquired during the cool-down. The magnetic field step occurs always at the SC transition temperature $T_{c}$, but it seems to be different from one series to another because of the not perfect thermal equilibrium between cavity and thermometers, especially for fast cool-downs from high temperatures.\\
\indent In FIG. \ref{fig:BB0}a the ratio between the magnetic field after and before the SC transition, acquired with the vertical flux gate at the mid position of the cavity, is shown as function of temperature. This ratio gives an idea of the magnetic field expelled after the SC transition, it is clear that for 4Ort the flux expulsion appears less efficient than for the other series, meaning more magnetic field remained trapped in the cavity wall, in agreement with the SC phase nucleation dynamics. In FIG. \ref{fig:BB0}b the ratio $B/B_{0}$, acquired with the vertical flux gate on top, as function of temperature gives instead informations regarding the magnetic field trapped on top. This data further corroborate the fact that the magnetic field for 4Ort is homogeneously trapped instead of being preferentially concentrated on top.\\
\indent One important point to notice is that the series 3Ax and 2Ort show the same $\Delta T_{bot-top}$ and $\Delta T_{mid-top}$ thermogradients (FIG. \ref{fig:DT}), and also the same SC-NC transition dynamics (FIG. \ref{fig:Cool}). This implies that the two cool-downs can be considered comparable, and the difference in residual $R_{0}$ by a factor of $2$ between them can be more likely attributed to the different orientation of the magnetic field.\\
\indent It is important to point out that the magnetic field just before the transition was slightly higher in case of the 2Ort (about $12$ mG) than in the case of 3Ax (about $9$ mG) yet this difference does not seem sufficient to account for a factor of two difference in residual resistance per same cooling regime. In general, looking at FIG. \ref{fig:QovsEacc} and at the residual resistances listed in TABLE \ref{table:table}, we have further hints that the magnetic field applied orthogonally to the cavity axis may have a larger effect in deteriorating the cavity performance and increasing the residual losses than the axial magnetic field. This possibility will be addressed more in detail in future studies with the use of temperature mapping, advance cavity diagnostic technique to study surface heating\cite{10}. \\
\indent Most importantly, data presented in FIG. \ref{fig:Temp} further corroborates the flux hole scenario. We discover that the orthogonal magnetic field can lead to local heating on top of the cavity equator, where we believe the magnetic field should concentrate after the SC transition. Indeed, during the acquisition of the series 1Ort, 2Ort and 3Ort the thermometers at the top position warmed up at high field, as shown in FIG. \ref{fig:Temp}. This temperature rising was prominent in the 1Ort series, where the temperature starts to exceed $1.4$ K at about $20$ MV/m, and it reaches $1.6$ K at about $30$ MV/m. The warming up is lower for the 2Ort and 3Ort series where it starts from about $27$ MV/m reaching just $1.45$ K. The absence of heating of 4Ort should be due to the different cool-down regime discussed previously, which does not involve the concentration of magnetic flux at the very top of the cavity, but rather flux being homogeneously trapped because of lack of cooling thermogradients.\\
\indent In the other cases instead, the heating on the very top position of the cavity is a newly described phenomenon for SRF cavities, and it can be considered a proof of the local dissipation due to concentrated trapped flux, pinned on top of the cavity when cooled in horizontal configuration, and in presence of orthogonal magnetic field.\\
\indent Interestingly, the same effect was repeatedly observed during test of the 9-cell nitrogen-doped niobium cavity TB9AES021 dressed with the LCLS-II vessel at the FNAL horizontal test facility (HTS). This cavity was instrumented with flux gates and thermometers inside the helium vessel, and as shown in FIG. \ref{fig:HTS}, the thermometer located on top of cell 1 of the cavity (input coupler side) showed significant heating starting at medium field (about $10$ MV/m) and the temperature reached values larger than $3$ K at above $20$ MV/m. This heating strongly affected cavity performance, causing $Q_{0}$ degradation.\\
\indent As it can be seen from FIG. \ref{fig:HTS}, increasing the starting temperature (cooling with larger thermogradients from $100$ K) pushed the onset of the heating and correspondently improved cavity performance. This nine cell data, together with the previously presented single cell data suggests a scenario where an orthogonal magnetic field component might be present close to cell 1 during the SC-NC transition, causing a "flux hole" hot spot to appear on top of cell 1, and suggest that this is an important performance limiting mechanism for superconducting cavities placed in an accelerator. Detailed results of a series of horizontal tests of nine cell N doped cavities dressed with different styles helium vessels will be presented in future works.
\section{\label{sec:level5}CONCLUSIONS}
This paper presents the first study of a superconducting single cell cavity cooled horizontally and immersed in different field orientations. The first important conclusion is that thermogradients at the SC-NC phase front may shrink towards the cavity top, causing more trapped flux close to an equatorial region, which translates into higher RF losses than in the case of vertical orientation. Cooling regimes solutions should therefore be sought in accelerators to ensure that a sufficient thermogradient is maintained throughout the full cell profile, or that the flux final resting place is not a cavity equator. Second important conclusion is that different field orientations may have a different impact on final performance; in particular, an orthogonal magnetic field may have a larger degrading impact for RF losses than an axial component, for same efficient cooling regime. Finally, an important new phenomena of heating at the top of the cavity has been observed in both the single cell and dressed nine cell studies, compatible with the "flux hole" scenario, where vertical field lines get encircled by superconducting regions and highly concentrated at the very top of the cavity. This can be harmful for cavity performance in an accelerator and could be leading to both Q-factor and quench degradation.\\
\section*{\label{sec:level6}ACKNOWLEDGEMENTS}
This work was supported by the US Department of Energy, Offices of High Energy Physics and Basic Energy Science, via the LCLS-II High Q Program. Authors would like to acknowledge technical assistance of A. Rowe, M. Merio, B. Golden, J. Rife, A. Diaz, D. Burk, B. Squires, G. Kirschbaum, D. Marks, and R. Ward for cavity preparation, testing and for cryogenics support. We acknowledge for fruitful discussions and support of the experiment M. Ross, R. Stanek and H. Padamsee. Fermilab is operated by Fermi Research Alliance, LLC under Contract No. DE-AC02-07CH11359 with the United States Department of Energy.
\section*{\label{sec:level7}REFERENCES}

\end{document}